\documentclass[twocolumn,showpacs,aps,amssymb,floatfix,prd,amsmath,preprintnumbers]{revtex4} 
\usepackage{epstopdf}
\usepackage{capt-of}
\usepackage{graphicx}  % Include figure files
\usepackage{dcolumn}   % Align table columns on decimal point
\usepackage{bm}% bold math
\begin{document}
%%%%%%%%%%%%%%%%%%%% Definitions %%%%%%%%%%%%

\def \reals{{\mathbb R}}
\def\be{\begin{equation}}
\def\ee{\end{equation}}
\def\bea{\begin{eqnarray}}
\def\eea{\end{eqnarray}}
\def\nn{\nonumber}
\def\th{\theta}
\def\ph{\phi}
\def\lt{\left}
\def\rt{\right}
\def\ed{\end{document}}
\def\degree{\mathop{\rm {{}^\circ}}}
\input epsf.tex
%%%%%%%%%%%%

%%%%%%%%%%%%
\title{Total Magnification and Magnification Centroid Due to Strongly Naked Singularity Lensing}

\author{Justin P.~DeAndrea and Shreya~Patel}
    %\email[]{}
     \affiliation{Department of Physics, The College of New Jersey, 2000 Pennington   
                   Rd., Ewing, NJ 08628, USA.}

%%%%%%%%%%%%
\begin{abstract}
A strongly naked singularity (SNS) was modelled at the center of the Galaxy. This specific type of naked singularity was characterized in 2008 by Virbhadra and Keeton. The lens is described using the Janis-Newman-Winicour metric, which has an ordinary mass and massless scalar charge parameters. Gravitational lensing by the SNS gives rise to 4 images: 2 images on the same side as the source and 2 images on the opposite side of the source from the optic axis. We compute magnification centroid, magnification centroid shift, and total absolute magnification for many values of the angular source position. The nature of the curve for all three results are qualitatively similar to Schwarzschild black hole lenses, but quantitatively different. Magnification centroid increases as angular source position increases. As angular source position increases, magnification centroid shift increases to a maximum value, and then begins to decrease. As angular source position becomes large, magnification centroid shift will approach zero. Total magnification is large for small values of angular source position, and decreases to a limiting value of one. The results expand upon previous work on an SNS of $\nu$ = 0.01. If such observations were viewed experimentally, they would show support for an SNS at the galactic center. Interpretation of an SNS would disprove the cosmic censorship hypothesis by Penrose, as well as the weaker cosmic censorship hypothesis by Virbhadra, which allows the existence of weakly naked singularities.

\end{abstract}

\pacs{95.30.Sf, 04.20.Dw, 04.70.Bw, 98.62.Sb }

%\keywords{Virbhadra-Ellis lens equation, black hole, and naked singularities}

\maketitle

%%%%%%%%%%%%%%%%%%%%%%%%%%%%%%%%%%%%%%%%%%

\section{Introduction}
A continuing problem in general relativity is proving or disproving the cosmic censorship hypothesis (CCH) developed by Penrose \cite{pen69,pen98,wald08}. This hypothesis states that a physically realistic gravitational collapse must end in the formation of a black hole (BH). Penrose postulates that a naked singularity cannot physically realistically form because the singularity would always be hidden within the event horizon of the BH \cite{pen98}. In recent years, many contradictions to the CCH have been developed \cite{pen98, prose, wald}. These contradictions generally fall into two major categories. The first category discusses how naked singularities are the end state of gravitational collapse; therefore, naked singularities violate the conditions sufficient for cosmic censorship. The second category contradicts the CCH by assuming the existence of naked singularities, and demonstrates that they are a violation of the CCH terms \cite{C2012}. \\
\indent
A major advance in strong field lensing involved the formulation of the Virbhadra-Ellis lens equation \cite{DeA, V2009}. The Virbhadra-Ellis lens equation theoretically models small and large deflections of light near a singularity, and generates image positions based on these conditions. This is important because large bending angles of light are commonly found in strong field lensing, and most equations cannot accurately model these large angles \cite{V2002}. Competing lens equations like the one proposed by Frittelli and Newman provide exact solutions \cite{F1999,F2000}. However, the results obtained are very similar to the results of the Virbhadra-Ellis lens equation. Therefore, the Virbhadra-Ellis lens equation is the preferred equation because it is much simpler and more commonly cited in the literature \cite{V2009}.\\
\indent
	There has been much research in recent years regarding gravitational lensing for various types of massive dark objects (MDO) \cite{P1,P2,P3,P4,P5,P6,P7,P8,P9,P10,P11,P12,P13,P14}. In order to probe the nature of the supermassive MDO in the center of the Milky Way, all research regarding lensing observables is very useful. The work of Virbradra et al., characterized 4 major types of singularities: the Schwarzschild black hole (SBH), weakly naked singularity (WNS), marginally strongly naked singularity (MSNS), and strongly naked singularity (SNS) \cite{V2002, V2008}. A SNS can be qualitatively distinguished from a SBH, WNS, and MSNS using strong field gravitational lensing. Additionally, qualitative differences were found amongst three types of SNS lensing, as previously described by Virbhadra and Keeton. These qualitative differences allowed Virbhadra and Keeton to describe 3 major types of SNS. The three major types are: Type 1 ($\nu$=0.04), Type 2 ($\nu$=0.02), and Type 3 ($\nu$=0.001) \cite{V2008}. \\
\indent
SNS can be qualitatively and quantitatively differentiated. SNS for $\nu$=0.04, $\nu$=0.02 and $\nu$=0.01 give rise to four images, and double Einstein rings when $\beta$=0 \cite{DeA, V2008}. SNS for $\nu$=0.001 give rise to four images, and no Einstein rings when $\beta$=0. Recently, DeAndrea studied a SNS between Type 2 and 3 ($\nu$=0.01) that interestingly found negative time delay results \cite{DeA}. Negative time delay signatures are only found for SNS; only positive time delay signatures are found for SBH, WNS, and MSNS \cite{DeA, V2008}. This paper expands upon the work of Virbhadra and Keeton, and considers an SNS with $\nu$=0.01. A model is proposed for a SNS as the Milky Way Galactic center. A static spherically symmetric Janis-Newman-Winicour (JNW) metric is used to determine light propagation \cite{jnw68}. We find this approach appropriate given the lack of scientific evidence in support of the CCH. \\
\indent
We compute magnification centroid, magnification centroid shift, and total absolute magnification for many values of $\beta$. Our results provide a deeper understanding of the lensing characteristics for the $\nu$=0.01 SNS. This is in addition to the time delay signatures DeAndrea previously explored \cite{DeA}. All computations were completed using MATHEMATICA software \cite{MATH}.\\

\section{\label{sec:LE&DE} Virbhadra-Ellis Lens Equation}

The Virbhadra-Ellis gravitational lens equation\cite{V2002} is presented here:
\begin{equation}
\tan\beta = \tan\theta - \alpha
\end{equation}
with
\begin{equation}
\alpha \equiv \frac{D_{ds}}{D_s} [\tan\theta + \tan(\hat{\alpha} - \theta)].
\end{equation}

$D_{s}$ refers to the observer-source distance, $D_{ds}$ the lens-source distance, and $D_d$ the observer-lens distance. $\alpha$ refers to the light bending angle. $\theta$  and $\beta$ refer to respectively, angular positions of an image and unlensed source measured from the optical axis. Here, the impact parameter is: $J = D_d sin \theta$. 

Refer to \cite{DeA} for a schematic diagram of gravitational lensing, showing all angles and distances presented in the lens equation. 

\section{\label{sec:LE&DE} Deflection Angle and Impact Parameter}

%%%%%%%%%%%%%%%%%%%%%%%%%%%%%%%%%%%%%%%%%%%%%%%%

Under the conditions of generally static and spherically symmetric spacetime, {\it Virbhadra et al.} \cite{vn98} characterized the line element as:
\begin{equation}
ds^2 = B(r)dt^2 - A(r)dr^2 - D(r)r^2(d\vartheta + sin^2\vartheta d\phi^2)
\end{equation}

For a light ray with the closest distance of approach $(r_0)$, the deflection $a(r_0)$ is characterized as \cite{vn98}:
\begin{equation}
\hat{\alpha}(r_0)=2\int_{(r_0)}^{\infty}\left(\frac{A(r)}{D(r)}\right)^\frac{1}{2} \left[\left(\frac{r}{r_0}\right)^2 \frac{D(r)}{D(r_0)}\frac{B(r_0}{B(r)} - 1\right]^\frac{1}{2}\frac{dr}{d} - \pi
\end{equation}
and the impact parameter $J(r_0)$ is characterized as \cite{vn98}:
\begin{equation}
J(r_0) = (r_0) \sqrt{\frac{D(r_0)}{B(r_0)}}.
\end{equation} 
\\
\indent
This is presented for a Schwarzschild SNS by \cite{VE2000}:

\begin{eqnarray}
\hat{\alpha}(x_0)=2\int_{x_0}^{\infty}\frac{dx}{\sqrt{(\frac{x}{x_0})^2\left(1-\frac{1}{x_0}\right)\left(1-\frac{1}{x}\right)}}-\pi
    \label{Schwarzschild strongly naked singularity} 
\end{eqnarray}
\\
where \cite{VE2000}:
\begin{eqnarray}
x = \frac{r}{2M} \ and \ x_0 = \frac{r_0}{2M}
    \label{Schwarzschild strongly naked singularity Second} 
\end{eqnarray}

\section{\label{sec:LE&DE} The Janis-Newman-Winicour Metric}

Janis, Newman and Winicour provided a general static and spherically symmetric solution to the Einstein massless scalar field equations. For constant and real parameters of the mass ($M$) and scalar charge ($q$), the line element (see \cite{jnw68,v97}) is characterized by: 
\begin{eqnarray}
   ds^2 &=&   \left(1-\frac{b}{r}\right)^{\nu} dt^2
      - \left(1-\frac{b}{r}\right)^{-\nu} dr^2 \nonumber \\
      &-& \left(1-\frac{b}{r}\right)^{1-\nu}  
      r^2 \left(d\vartheta^2  +\sin^2\vartheta \  d\varphi^2\right) 
    \label{Janis-Newman-WinicourMetric} 
\end{eqnarray}
The massless scalar field is characterized by:
\begin{equation}
\Phi = \frac{q}{b\sqrt{4\pi}}\ln\left(1-\frac{b}{r}\right),
\end{equation}
with:
\begin{equation}
\nu = \frac{2M}{b}\; \mbox {and}\;  b = 2\sqrt{M^2 + q^2}.
\end{equation}
There is only one photon sphere situated at the radial distance \cite{V2002,v97,vec01}:
\begin{equation}
r_{ps}=\frac{b(1+2\nu)}{2}.
\end{equation}
There  is a naked singularity at $r=b$. Only for $\frac{1}{2}<\nu\leq1$ does the photon sphere exist. Defining \cite{vjj}:
\begin{equation}
\rho=\frac{r}{b} \; \mbox{and} \; \rho_0=\frac{r_0}{b},
\end{equation}
the deflection angle $\hat{\alpha}$ for a light ray is \cite{V2002,vn98}:
\begin{widetext}
\begin{equation}
\hat{\alpha}(\rho_0)=2\int_{\rho_0}^{\infty}\frac{d\rho}{\rho\sqrt{1-\frac{1}{\rho}}\sqrt{(\frac{\rho}{\rho_0})^2\left(1-\frac{1}{\rho}\right)^{1-2\nu}\left(1-\frac{1}{\rho_0}\right)^{2\nu-1}-1}}-\pi.
\end{equation}
\vspace{2ex}
\end{widetext}

%%%%%%%%%%%%%%%%%%%%%%%%%%%%%%%%%%%%%%%%%%%%%%%%%%%%%%%%%%%% New Section: Magnification Centroid and Total Magnification Equations %%%%%%%%%%%%%%%%%%%%%%%%%%%%%%%%%%%%%%%%%%%%%%%%%%%%%%%%%%%%%%%%%%%%%%%

\section{Magnification Centroid and Total Magnification Equations}
The magnification centroid of images is characterized by:
%\begin{widetext}
\begin{equation}
\hat{\Theta}=\frac{\Sigma \theta |\mu_i|}{\Sigma |\mu_i|}
\end{equation}
%\end{widetext}
where angles measured clockwise to the optic axis are positive and angles counterclockwise are negative.
\\
\\
The magnification centroid shift is defined as:
\begin{equation}
\Delta\Theta=\beta - \hat{\Theta}
\end{equation}
The total magnification is defined as:
%\begin{widetext}
\begin{equation}
\mu_{tot}=\Sigma |\mu_i|
\end{equation}
%\end{widetext}

\section{\label{sec:LE&DE} Computations and Results}

Virbhadra and Keeton found several important trends in magnification and magnification centroid for SNS lensing. These trends were derived from extensive data on the Type 1 ($\nu$=0.04), Type 2 ($\nu$=0.02), and Type 3 ($\nu$=0.001) SNS. In our current article, we examine a $\nu$=0.01 case in order to distinguish lensing behavior of SNS between types 2 and 3. The supermassive center of the Milky Way galaxy is modelled as a SNS ($\nu$ = 0.01, mass $M$ = 3.61 x $10^6 M_{\bigodot}$, $D_d$ = 7.62 kpc, $D_{ds}$/$D_s$=$\frac{1}{2}$). \\
\indent
We calculate magnification centroid, magnification centroid shift, and total magnification using MATHEMATICA, for a large number of angular source positions ($\beta$). The Galactic MDO is modelled as a JNW SNS lens, as the JNW spacetime is proven stable under scalar field perturbations \cite{Sadu}. Due to the nature of SNS having no photon sphere, there should be no relativistic images produced, thus we did not attempt to study relativistic images in this paper \cite{V2002}. \\
\indent
Table I displays values of magnification centroid, magnification centroid shift, and total magnification for a large number of angular source positions. Refer to Tables 1-2 by \cite{DeA} for the quantitative values of image positions for a large range of $\beta$. Figures 1-5 are schematic diagrams of SNS gravitational lensing, visualizing image positions as $\beta$ increases. $I_{oo}$ denotes the outer image on the opposite side of the source, $I_{io}$ is the inner image on the opposite side of the source, $I_{is}$ is the inner image on the same side of source, and $I_{os}$ is the outer image on the same side of source. Figure 1 shows SNS gravitational lensing at $\beta$=0, where a double Einstein ring is present. The schematic diagrams denoted Figure 2 and Figure 3 show image positions as $\beta$ increases. Note that as $\beta$ increases, $I_{oo}$ and $I_{io}$ come closer, while $I_{is}$ and $I_{os}$ separate. Figure 3 represents the point where the light ray of $I_{is}$ will invert its deflection. Figure 4 shows a radial caustic, when $I_{io}$ and $I_{oo}$ are located in the same position, and become highly magnified. As $\beta$ further increases, $I_{io}$ and $I_{oo}$ will eventually disappear, shown in Figure 5. The nature of curve for Figures 6-8 are all qualitatively similar to black holes, but quantitatively different. The quantitative differences can be used to classify the type of MDO lens.\\
\indent
 Magnification centroid for given values of $\beta$ follows a roughly linear trend, with slight deviations initially observed for small values of $\beta$. Magnification centroid shift is zero when $\beta$=0. At $\beta$=0 two Einstein rings can be observed. Magnification centroid shift increases to a maximum value as $\beta$ approaches 2 arcseconds, then continues to decrease to a limiting value of 0. The total magnification is very large initially, and as $\beta$ increases the total magnification decreases to a limit of 1.\\

\begin{center}
\includegraphics[width=3in,height=3.37in]{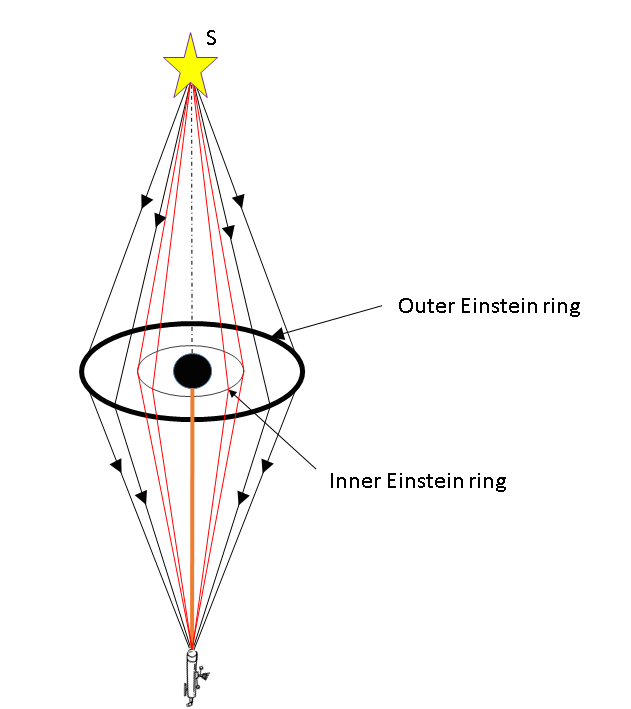}
\DeclareGraphicsExtensions{.png}
\captionof{figure}{SNS Gravitational lensing at $\beta$􏰂=0, showing a double Einstein ring}
\label{FIGURE1} % 
\end{center}

\begin{center}
\includegraphics[width=3in,height=3.68in]{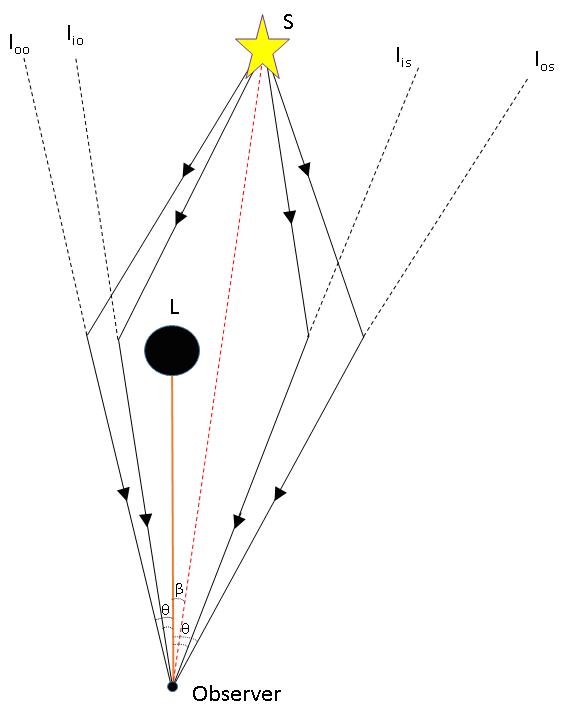}
\DeclareGraphicsExtensions{.png}
\captionof{figure}{Schematic diagram of gravitational lensing by a SNS}
\label{FIGURE2} % 
\end{center}

\begin{center}
\includegraphics[width=3in,height=3.68in]{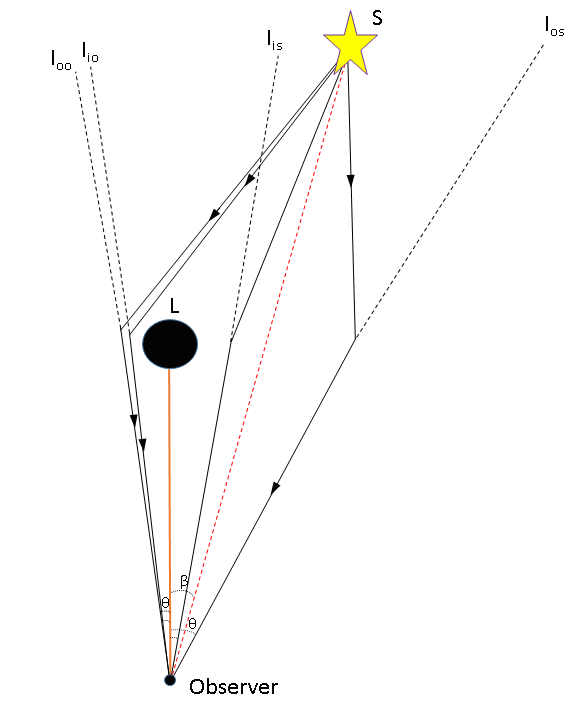}
\DeclareGraphicsExtensions{.png}
\captionof{figure}{Schematic diagram of gravitational lensing by a SNS}
\label{FIGURE3} % 
\end{center}

\begin{center}
\includegraphics[width=3in,height=3.68in]{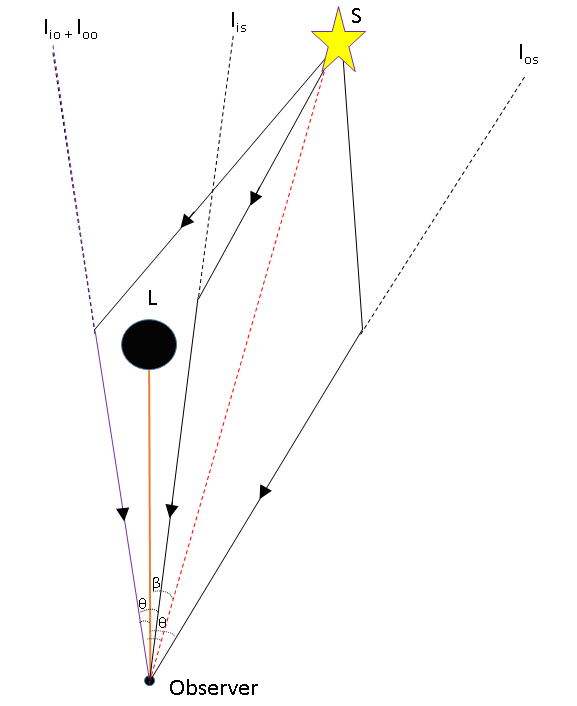}
\DeclareGraphicsExtensions{.png}
\captionof{figure}{Schematic diagram of gravitational lensing by a SNS, where values of 􏰂$\beta$ for secondary images coincide}
\label{FIGURE4} % 
\vspace{15ex}
\end{center}

\begin{center}
\includegraphics[width=3in,height=3.68in]{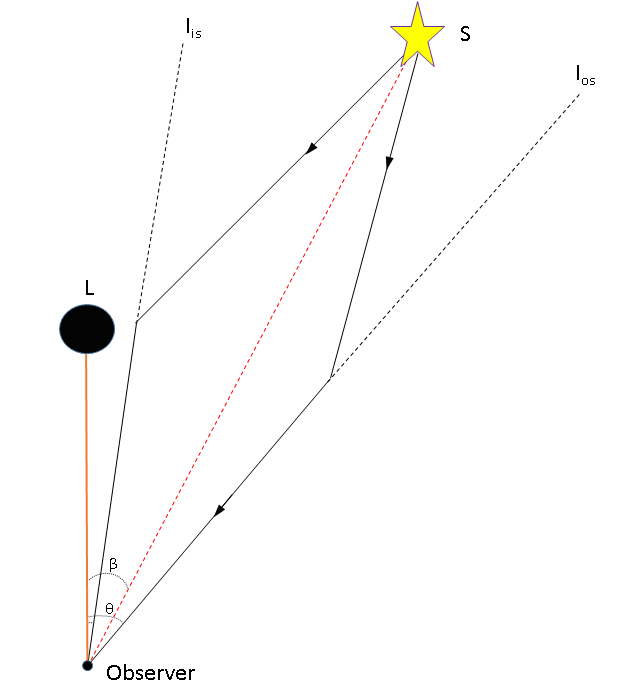}
\DeclareGraphicsExtensions{.png}
\captionof{figure}{Schematic diagram of gravitational lensing by a SNS, where both secondary images are annihilated}
\label{FIGURE5} % 
\end{center}

\begingroup
\squeezetable
\begin{table*}
\caption{\label{tab:Table1} We model the Galactic supermassive center as a strongly naked singularity ($\nu$ = 0.01, mass $M$ = 3.61 x $10^6 M_{\bigodot}$, $D_d$ = 7.62 kpc, $D_{ds}$/$D_s$=$\frac{1}{2}$) and, using Mathematica, compute magnification centroid, magnification centroid shift, and total magnification for a large number of angular source positions ($\beta$).}
          
%%%%%%%%
\begin{ruledtabular}
 \begin{tabular}{l|lllllll}
%%%%%%%%
\multicolumn{1}{c|}{ $\beta (arcsec)$}&
\multicolumn{6}{c}{Images on the opposite side of the source.}\\
%%%%%%%%% 
  &         $\hat{\Theta} (arcsec)$                &    $\Delta\hat{\Theta} (arcsec)$                   &   $\mu_{tot}$                  \\
\hline
 \hline\noalign{\smallskip}
%%%%%%%%%%%%%%%%%%%%%%%%%%%%%%%%%%%%%% 
$10^{-6}          $&$1.5 * 10^{-6}            $&$5.00 * 10^{-7}          $&$ 1.3882 * 10^{6} $ \\     
$10^{-5}    $&$1.5 * 10^{-5}            $&$5.00 * 10^{-6}          $&$ 1.3882 * 10^{5}  $ \\      
$10^{-4}    $&$1.5 * 10^{-4}            $&$5.00 * 10^{-5}          $&$ 1.3882 * 10^{4} $ \\      
$10^{-3}    $&$1.5 * 10^{-3}            $&$5.00 * 10^{-4}          $&$ 1.3882 * 10^{3}$   \\       
$10^{-2}          $&$1.5 * 10^{-2}            $&$5.00 * 10^{-3}          $&$ 1.3882 * 10^{2}$   \\       
$10^{-1}         $&$1.5 * 10^{-1}            $&$4.99 * 10^{-2}          $&$ 1.3909 * 10^{1} $  \\     
$1.0         $&$1.40            $&$0.397          $&$ 1.6449$ \\         
$2.0        $&$2.49            $&$0.491          $&$ 1.1477 $ \\          
$3.0        $&$3.45           $&$0.450          $&$ 1.0482$\\           
$4.0        $&$4.39           $&$0.388          $&$  1.0194$ \\           
$5.0        $&$5.33           $&$0.334          $&$ 1.0090  $ \\           
$6.0       $&$6.29            $&$0.290         $&$  1.0047$ \\          
$7.0          $&$7.26         $&$0.255          $&$ 1.0027$ \\           
$8.0   $&$8.23               $&$0.227          $&$1.0016  $ \\         
$9.0   $&$9.20               $&$0.204          $&$1.0010$ \\         
$10.0   $&$10.2               $&$0.186          $&$1.0007   $ \\         

%%%%%%%%%%%%%%%%%%%%%%%%%%%%%%%%%
\end{tabular}
\end{ruledtabular}
\vspace{10ex}
\end{table*}
\endgroup

\begin{center}
\includegraphics[width=3in,height=2.05in]{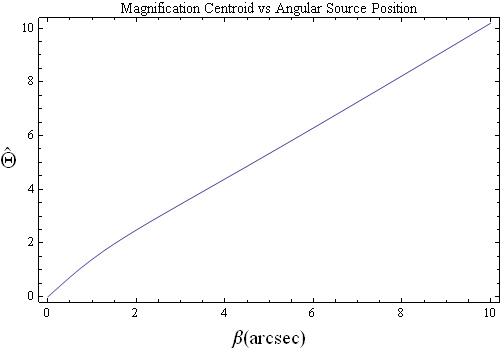}
\DeclareGraphicsExtensions{.png}
\captionof{figure}{The magnification centroid shift 􏰀$\hat{\Theta}$ plotted against angular source position $\beta$ for  $\nu$ = 0.01; $D_{ds}/D_s$=$\frac{1}{2}$}
\label{Fig1} % 
\end{center}

%%%%%%%%%%%%%%Figure 2 %%%%%%%%%%%%%%%%%%%%%%%%%%
\begin{center}
\includegraphics[width=3in,height=2.05in]{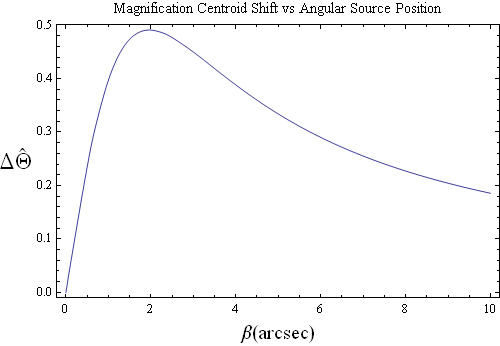}
\DeclareGraphicsExtensions{.png}
\captionof{figure}{The magnification centroid shift 􏰀$\Delta$$\hat{\Theta}$ plotted against angular source position $\beta$ for  $\nu$ = 0.01; $D_{ds}/D_s$=$\frac{1}{2}$}
\label{Fig2} % 
\end{center}

%%%%%%%%%%%%%%Figure 3 %%%%%%%%%%%%%%%%%%%%%%%%%%
\begin{center}
\includegraphics[width=3in,height=2.05in]{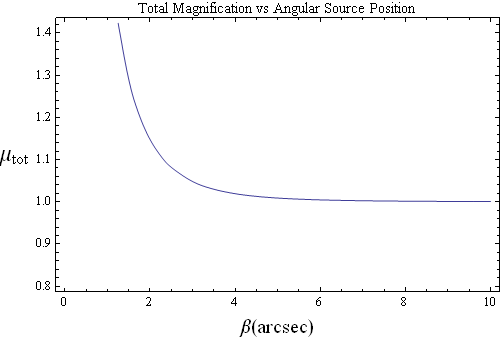}
\DeclareGraphicsExtensions{.png}
\captionof{figure}{Total magnification (􏰀$\mu_{tot}$) plotted against angular source position $\beta$ for  $\nu$ = 0.01; $D_{ds}/D_s$=$\frac{1}{2}$}
\label{Fig3} % 
\end{center}

%%%%%%%%%%%%%%%%%%%%%%%%%%%%%%%%%%%%%%%%%%%%%%%%%%%%%%%%%%%%

\section{Summary and Discussion}

We are able to model the magnification centroid, magnification centroid shift, and total absolute magnification for a strongly naked singularity with $\nu$=0.01. When images are not resolved, total magnification and magnification centroid contribute greatly to our study of lensing. These results demonstrate that an SNS with $\nu$=0.01 has quantitatively different values for all three parameters compared to SNS with other values of $\nu$. Therefore, the quantitative values of magnification centroid, magnification centroid shift, and total absolute magnification depend on the value of $\nu$=0.01 even among SNS \cite{V2008}. Compared to the SBH, the magnification centroid will be quantitatively lower and total magnification will be quantitatively higher for a constant value of $\beta$ \cite{V2002, V2008}. This is due to the higher value of the scalar charge to mass ratio. \\
\indent
However, our results for $\nu$=0.01 SNS show qualitatively similar trends to the SNS, SBH, WNS, and MSNS trends that were previously described by Virbhadra and Keeton \cite{V2008}. Despite the differences in $\nu$ value, the four types of singularities show similar trends. It is found that as $\beta$ increases, magnification centroid increases as long as $\nu$ is constant. In addition, as $\nu$ decreases, magnification centroid decreases for a constant value of $\beta$. \\
\indent
The magnification centroid shift, which is mathematically derived from the magnification centroid, also shows similar trends based on $\beta$ values; this is for SNS, SBH, WNS, and MSNS. Magnification centroid shift rapidly increases with $\beta$, attains a peak value, and then slowly declines to 0. Finally, when ν is constant, total magnification has a high value for very small values of $\beta$. This value of total magnification rapidly decreases as $\beta$ increases. When $\beta$ is constant, as $\nu$ increases, total magnification also increases. This increase in total magnification is likely due to the scalar charge. \\
\indent
The values for magnification centroid, magnification centroid shift, and total absolute magnification add to the current knowledge on SNS with $\nu$=0.01. Previously, DeAndrea explored time delays for SNS lensing with $\nu$=0.01. It was found that for the $\nu$=0.01 case, unlike the types of SNS studied by Virbhadra and Keeton, all four images have negative time delays even for $\beta$=0. The primary image has a negative time delay for any value of $\beta$, and the time delays for the other three images remained negative until very large values of $\beta$ \cite{DeA}. This was not found in the results of Virbhadra and Keeton \cite{V2008}. \\
\indent
Observations of our results would disprove the Cosmic Censorship Hypothesis and would pave the way to develop a quantum theory of gravity, as very strong gravitational fields will be accessible to observation. Next generation imaging technology could attain the goal of experimentally imaging a black hole \cite{K2005}. The NASA MAXIM mission plans to create an X-ray interferometer with 100 nanoarcsecond resolution \cite{NASA}. Since imaging a black hole is attainable at at 300 nanoarcsecond resolution, this project makes it more of a reality. To minimize demagnification; the source, lens, and observer must be highly aligned. When making observations of the universe at great distances, uncertainty will be present due to cosmic variance \cite{B2000}. Since we considered the Janis-Newman-Winicour spacetime with a scalar field, we must expect effects on lensing due to dark energy \cite{H2013}. Thus, these results could be potentially useful for the study of dark energy lensing. The absorption and scattering of electromagnetic radiation near the galactic center causes additional difficulties. 

\section{Acknowledgements}

We thank our mentor K. S. Virbhadra for the time he took to contribute to our learning. His enthusiasm for this field motivated us to reach new levels of understanding and intelligence in astrophysics. All in all, we appreciate everything that he has done for us the last several years.

\end{document}